# Investigating Speed Deviation Patterns During Glucose Episodes: A Quantile Regression Approach

Aparna Joshi[1], Jennifer Merickel[2], Cyrus V. Desouza[2], Matthew Rizzo[2], Pujitha Gunaratne[3], and Anuj Sharma[1]

**Abstract—** Given the growing prevalence of diabetes, there has been significant interest in determining how diabetes affects instrumental daily functions, like driving. Complication of glucose control in diabetes includes hypoglycemic and hyperglycemic episodes, which may impair cognitive and psychomotor functions needed for safe driving. The goal of this paper was to determine patterns of diabetes speed behavior during acute glucose to drivers with diabetes who were euglycemic or control drivers without diabetes in a naturalistic driving environment. By employing distribution-based analytic methods which capture distribution patterns, our study advances prior literature that has focused on conventional approach of average speed to explore speed deviation patterns.

## I. INTRODUCTION

Given the global prevalence of diabetes [1], there have been ongoing and important research developments about how diabetes impacts instrumental daily activities, including driving. Drivers with diabetes may have an elevated risk of being involved in traffic accidents [2]. Hypoglycemia, prevalent in type 1 diabetes (T1DM) increases this risk and may reduce driver compliance to roadway speed limit ([3], [4]). The American Diabetes Association recommends that clinicians consider this factor when advising patients on driver safety [5].

While hypoglycemia and driving have received significant research attention ([6], [7], [8]), less research has focused on hyperglycemic impacts on driving [9], [10]. Hyperglycemia may also impair abilities needed for safe driving ([11], [12]), is a prevalent diabetes complication [13], and may particularly impact driving on higher speed roadways suggesting increased risk [14]. Hyperglycemia is particularly prevalent in type 2 diabetes (T2DM) and some research has suggested that hyperglycemia may impact unsafe stopping behavior more than hypoglycemia [15]. This highlights the need to improve understanding of how acute hyperglycemia impacts driver safety in diabetes across real-world context.

Our goal was to improve understanding of how acute hyperglycemia, in the context of individual variation, affects diabetes speed control. This research fits within the broader goals of developing supportive in-vehicle technology to support drivers with diabetes. To meet this goal, we investigated drivers' deviation from speed limit patterns linked to acute hyperglycemia in T1DM and T2DM using quantile regression.

Speed control was selected due to its association with fatal crashes [16]. Quantile regression advances prior work by going beyond population averages, to provide a more comprehensive examination of the entire speed control distribution. This analysis is focused on uninterrupted flow roadways, like interstates.

## II. DATA COLLECTION

Data for this study was taken from registries at the University of Nebraska Medical Center (UNMC). Participants consented to data collection and registry participation following UNMC IRB guidelines (IRBs #208-18-FB and #462-16-FB).

**Drivers:** Participants with type 1 diabetes mellitus (T1DM: N=17, male=7), type 2 diabetes mellitus (T2DM: N=62, male=36), and controls without diabetes (N=58) were included (Table I). Miles driven and the number of drives for each participant group is provided in Table I. Drivers were screened for eligibility at study start. Presence or absence of diabetes was confirmed based on HbA1c blood labs (T1DM: <12% HbA1c; T2DM: <7.5% HbA1c; Controls: <5.7% HbAlc), clinical exam, and diagnosis history. Clinical exams were conducted by an experienced endocrinologist.

TABLE I. SUMMARY OF STUDY PARTICIPANTS

| Variable | Disease Type 1 (N=31) | | Disease Type 2 (N=106) | |
|---|---|---|---|---|
| | *T1DM* | *Control-1* | *T2DM* | *Control-2* |
| **Age (years)** | | | | |
| Mean (SD) | 30.4 (7.9) | 38.6 (10.7) | 60.5 (5.9) | 59.7 (8.0) |
| Range | 21-52 | 21-55 | 44-70 | 39-70 |
| **Sex** | | | | |
| Female (%) | 10 (58.8%) | 10 (71.4%) | 26 (41.9%) | 18 (40.9%) |
| Male (%) | 7 (41.1%) | 4 (28.5%) | 36 (58.0%) | 26 (59.0%) |
| **Total Participants** | 17 | 14 | 62 | 44 |
| **Miles Driven** | 16233 | 17752 | 57385 | 88158 |
| **Number of Drives** | 1999 | 1634 | 12367 | 7652 |

All drivers were legally licensed, experienced, active drivers. All participants met Nebraska Department of Motor Vehicle (DMV) standards (visual acuity of <20/50 OU

\* Research supported by Toyota Collaborative Safety Research Center.

1 Aparna Joshi and Anuj Sharma are with Institute for Transportation, Iowa State University, Ames, Iowa, USA (phone: 515-817-7942; e-mail: aparnaj8@iastate.edu).

2 Jennifer Merickel, Cyrus V. Desouza, and Matthew Rizzo are with University of Nebraska Medical Center, Omaha, Nebraska, USA.

3 Pujitha Gunaratne is with the Toyota Collaborative Safety Research Center, Toyota Motor North America, Ann Arbor, USA.

corrected or uncorrected). All drivers had no significant, confounding medical conditions (e.g., peripheral nerve, renal, neurological, cardiovascular, or major psychiatric diseases; significant mobility impairment; substance use in the past year) or medication usage (e.g., narcotics, sedating antihistamines, and major psychoactive medication). Eye diseases (e.g., retinopathy) were permitted in drivers with diabetes if they met Nebraska state licensure standards for vision. Control drivers were matched drivers with diabetes based on age (within 5 years), sex, season of driving in the study (winter or not winter), and similar medical history (except for diabetes). T1DM drivers used insulin at least daily. T2DM drivers were on insulin, a sulphonylurea drug, or any two diabetic drug combinations.

**NDD:** All drivers had a sensor system installed into their personal vehicle for the duration of the study. T1DM drivers participated for 4-weeks and T2DM drivers participated for 6-weeks. Drivers drove as typical around Omaha, NE, and surrounding areas during the study. Driving data were collected at a frequency of every second from on- to off-ignition. The collected data included video recordings of the forward roadway and the vehicle cabin, as well as vehicle sensor data such as GPS, speed, and accelerometer.

**CGM Data:** Glucose levels were monitored throughout study participation in drivers with diabetes using the FDA approved Dexcom G4 Platinum Professional continuous glucose monitor (CGM). CGMs were blinded (the drivers could not see their glucose levels) and sampled glucose levels every 5 minutes. Drivers with diabetes could not be using a CGM prior to study participation. CGM data were post-processed for quality according to FDA standards [17]. Drivers with diabetes calibrated CGMs twice daily by entering self-sampled blood glucose readings from their blood glucose meter.

**Acute Glucose Episodes:** Acute glucose levels were labeled as diabetic hyperglycemia (DHR; >180 mg/dL) and euglycemia (DN; 71-179 mg/dL) according to the American with Diabetes Association guidelines [18]. Control data were labeled as HC.

### III. DATA PROCESSING

**Data were cleaned** to remove missing data. Missing GPS and speed data represented 4% of total driving data in T1DM and T2DM datasets. Missing CGM data accounted for 5.5% (T1DM) as documented in reference [15], and an analogous approach was undertaken when addressing missing CGM data for T2DM.

**Vehicle speed data were** aggregated into 45-second time chunks, from start to end of drive (N =113397). This process summarized the data within each time chunk (e.g., average vehicle speed), and reduced noise from second-by-second driver speed variation.

**Speed limit data were** extracted by mapping vehicle GPS data to GIS databases for Nebraska from the US TIGER Road files and Here Maps databases. We retrieved posted speed limits (PSLs) spanning 10 mph to 75 mph. Next, the data were filtered to only include PSLs 75, 70, and 65 mph, to study deviation from speed limits on interstate segments. This included 34% and 27.4% of data available in Nebraska.

To ensure adequate sample sizes for analysis across the DHR, DN, and HC groups, PSLs for 70 mph were removed from further analysis due to insufficient hyperglycemia observations (N=45, 5.8% of 70 mph data) (see Table II for a listing of data across three distinct speed limits).

TABLE II. DATA ACROSS SELECTED SPEED LIMITS

| Disease Type | Glucose Episode | Speed Limits (mph) | | |
|---|---|---|---|---|
| | | 75 | 70 | 65 |
| T1DM | Control | 1223 | 535[a] | 2715 |
| | Normal | 794 | 181[a] | 1322 |
| | Hyperglycemic | 257 | 45[a] | 402 |
| | Total Participants | 12 | 21[a] | 30 |
| | Road Segments | 25 | 46[a] | 28 |
| T2DM | Control | 1111 | 1250 | 10923 |
| | Normal | 1378 | 1858 | 6861 |
| | Hyperglycemic | 257 | 490 | 611 |
| | Total Participants | 23 | 78 | 101 |
| | Road Segments | 26 | 110 | 354 |

[a.] Not selected in data analysis.

### IV. DATA ANALYSIS

**The deviation from speed limit outcome** was calculated as the difference between the vehicle's speed and the posted roadway speed limit for each second of the drive and then averaged for each 45-second time chunk. Higher values indicate driver speeding.

**The primary predictors** were either the driver with diabetes **glucose status** (DHR or DN) or the **disease status** of the driver (DN or HC).

**Traffic flow period control covariates** captured **day of the week** (weekday vs. weekend) and **time of day** (peak vs. off-peak hours) based on typical traffic flow. For weekday drives, the peak hours are defined as the period between 6:00 AM and 9:30 AM, and between 2:00 PM and 6:30 PM. On weekends, the peak hours are classified as the period between 11:00 AM and 6:00 PM. All other drives outside of these specified peak hours are considered off-peak hours. The impact of these temporal elements as well as sex, on speed selection has been a subject of prior investigation ([19]-[23]). Heightened traffic flow during peak hours, and potentially during shoulder weekdays, elevated the probability of adhering to speed limits ([19], [20]).

**Sex** was another control covariate included to account for males showing typically higher tendency to speed compared to females ([21]-[23]).

**Analytical Method 1:** Initially, we used the non-parametric **Kolmogorov-Smirnov Test (KS test)** to assess if differences in deviation from speed limit were seen across the selected PSLs. The KS test was selected because it makes no assumption regarding the data's distributional properties.

**Analytical method 2: Quantile Regression (QR)** was used for the inspection of changes associated with acute glucose or driver disease across the entire driver's deviation from speed limit distribution. QR approaches do not assume a distributional shape to the outcome, allowing robustness to data extremes and outliers. QR explores how the relationship between *x* and *y* varies across different points of the distribution, providing a full picture of the entire conditional distribution of *y* [24]. Equation (1) explains QR.

$$y_i = x_i \beta_q + e_i \quad (1)$$

where, $\beta_q$ is the vector of unknown parameters associated with the $q^{th}$ quantile.

We modeled the five quantiles (0.25, 0.50, 0.75, 0.85, 0.90) across the drivers' deviation from speed limit stratified by PSLs (dependent variable). In each PSL, we assessed if acute glucose status or disease (DN, DHR, HC) impacted deviation from speed limit, while controlling day of week, time of day, and sex. For QR modeling, we performed one-hot encoding to treat categorical glucose variables and covariates (Table III).

TABLE III. VARIABLE DESCRIPTION FOR ONE-HOT ENCODING

| Variable | Variable Type | Variable Levels | | |
|---|---|---|---|---|
| Deviation from Speed limit | Continuous | - | - | - |
| Glucose Episode | Categorical | DN[b] | HC | DHR |
| Traffic Flow Period | Categorical | Off-peak hours[b] | Peak hours | |
| Type of Day | Categorical | Weekend[b] | Weekday | |
| Sex | Categorical | Male[b] | Female | |

[b.] Reference category selected for one-hot encoding.

## V. RESULTS

### A. Cumulative Distribution Plots and KS Test Results

The KS test results are shown in Table IV. Except for 65 mph in T1DM, the KS test confirms that there were significant differences in speed limit adherence associated with acute glucose status (DN vs DHR) and disease status (DN vs HC) in T1DM and T2DM. Since the p values are less than 0.05, we conclude that there are significant differences in terms of speed deviation distribution between DHR vs DN and HC vs DN in T1DM and T2DM. Fig. 1 (T1DM) and Fig. 2 (T2DM) shows the overlaid cumulative distribution plots (CDFs) for deviation from speed limit across the PSLs 75, 70, and 65 mph. The distribution of data for 65 mph in T1DM shows an overlap between DHR and DN, consistent with the findings of the KS test.

TABLE IV. KOLMOGROV-SMIRNOV TEST RESULTS

| Category | Speed Limit (mph) | KS Stat (p-value) | |
|---|---|---|---|
| | | Normal vs Control | Normal vs Hyperglycemia |
| T1DM | 75 | 0.432(<0.05) | 0.389(<0.05) |
| | 65 | 0.313(<0.05) | 0.068(0.104) [c] |
| T2DM | 75 | 0.429(<0.05) | 0.299(<0.05) |
| | 70 | 0.244(<0.05) | 0.186(<0.05) |
| | 65 | 0.186(<0.05) | 0.172(<0.05) |

[c.] Not significant at 95% level of confidence.

### B. Quantile Regression (QR) Results

Since speed deviation metric showed discrimination power in KS test results, next we performed a QR to understand the difference in DHR vs DN and HC vs DN on deviation from speed limit at different quantiles. Table V shows the estimated parameters and their p-value for T1DM and T2DM models. While most of the coefficients are significant at 99% confidence level, coefficients at 95% confidence level are also reported. The coefficients were identified relative to specific reference categories as mentioned in Table III.

**Quantile Regression Results for T1DM Drivers**

- Deviation from speed limit during DHR was statistically significant in most of the quantiles except for 85[th] (at 75 mph) and 50[th] and 75[th] (at 65 mph) in T1DM.

- Compared to DN, DHR in T1DM contributed to the highest positive deviation (2.6 mph) from 75 mph in the lower portion of the distribution. Contrastingly, higher quantiles showed higher deviating speed from 65 mph for DHR.

- On comparing HC drivers to T1DM drivers, a noticeable distinction can be seen for 75 mph and 65 mph. While HC drivers had operating speeds above 75 mph throughout the distribution, HC at 65 mph traveled at speeds below the speed limit.

- **Traffic Flow Period:** The insignificance of the traffic flow period was predominantly observed within the middle and higher quantiles for both the 75 mph and 65 mph PSLs.

- **Type of Day:** Speed distribution at 75 mph revealed that weekday drives had lower deviation from speed limit as compared to weekends across all quantiles. Contrasting results are seen for 65 mph PSL.

- **Sex:** A higher deviation from speed limit was observed for females than males at 75 mph for all quantiles except for 25[th]. In contrast, differing outcomes were observed for the 65mph posted speed limit where females showed lower deviation than their male counterparts.

Although our hypothesis regarding the impact of glucose episodes on deviation from speed limit aligns with the previous study [25], our investigation reveals contrasting trends for hyperglycemia. Unlike the previous study, which observed negative deviation from speed limit during hyperglycemia episodes among individuals with T1DM, our findings differ in this regard [25]. We also observed insignificant differences for 65 mph (25th, 50th, 75th quantiles) and 75 mph (85th quantile). A significant factor contributing to this difference may stem from their utilization of a linear mixed effects model (LMM), which estimated the mean deviation from speed limit during hyperglycemia. Furthermore, their negative deviation from speed limit during hyperglycemia within T1DM were for speed limits above 15 mph in general. We suggest that the relationship between glucose episodes and deviation from speed limit may vary depending on the speed limit range being considered.

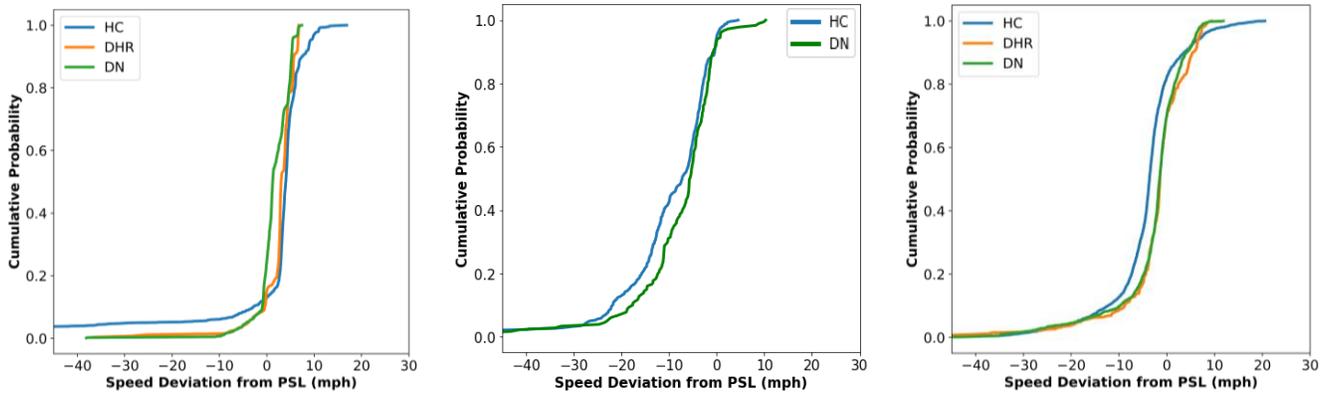

Fig. 1. 75 mph (left), 70 mph (middle), 65 mph (right) Cumulative Distribution Plots for T1DM Disease Type

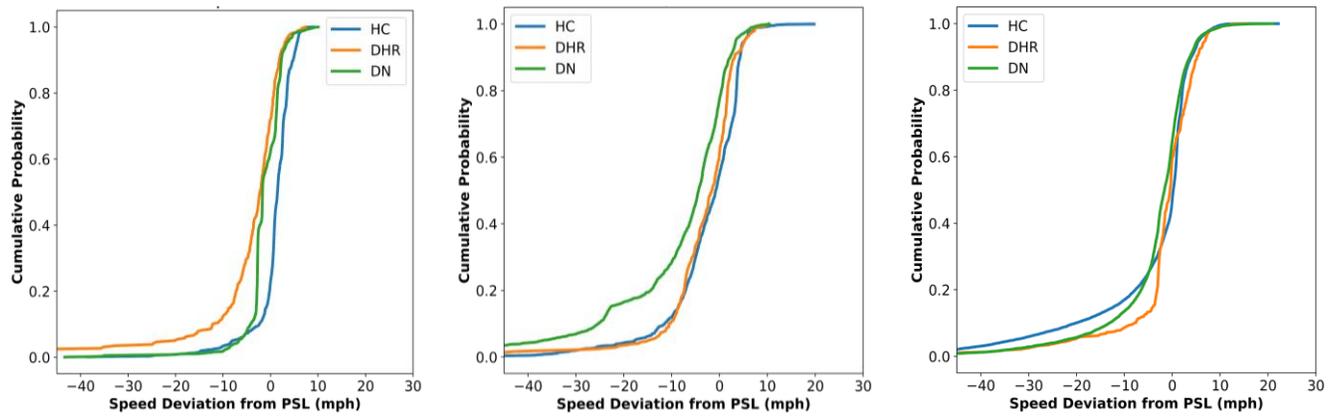

Fig. 2. 75 mph (left), 70 mph (middle), 65 mph (right) Cumulative Distribution Plots for T2DM Disease Type

**Quantile Regression Results for T2DM Drivers**

- In T2DM, all the quantiles within every PSL indicated significant differences in DHR and DN speed limit adherence behavior.

- Against DN group, the DHR group traveled below the 75mph speed at $25^{th}$, $50^{th}$ and $75^{th}$ quantile, in contrast to $85^{th}$ and $90^{th}$ percentile where the deviation from speed limit was positive.

- The speed distribution at 70 mph showed that participants during DHR have deviation from speed limit higher than DN drivers, and the increase was higher in the lower portion of the distribution.

- A similar speed behavior during hyperglycemia was seen for 65 mph and the highest speed variation occurred at $85^{th}$ percentile (1.778 mph).

- Additionally, a positive variation from speed limit was noticed for HC drivers in all the quantiles of 75 mph, 70 mph, and 65 mph, except for the $90^{th}$ quantile of 65 mph in T2DM.

- **Traffic Flow Period:** At 75 mph, deviation from speed limit experienced a decrease during peak hours compared to off-peak hours. A parallel pattern emerged at the 25th, 75th, and 85th quantiles for 70 mph, as well as at the 25th and 50th quantiles for 65 mph.

- **Type of Day:** Deviation from speed limit distribution at 75 mph showed a propensity for individuals to display elevated deviation from speed limit on weekdays than weekends, across all quantiles except for $25^{th}$. Conversely, the examination of the 70 mph and 65 mph yields divergent findings.

- **Sex:** Females showed a higher degree of deviation from speed limit than males at 75 mph, 70 mph, and 65 mph.

**Quantile Regression Results for T1DM vs T2DM Drivers**

- Additionally, a positive variation from speed limit is noticed for control group (HC) in all the quantiles of 75 mph, 70 mph, and 65 mph, except for the $90^{th}$ quantile of 65 mph in T2DM.

- Like T2DM drivers, T1DM drivers during hyperglycemic episodes (DHR) showed higher deviation from speed limit from regulatory speed limits (75 mph and 65 mph) as compared to their euglycemic episode (DN) counterparts. However, this observation was reversed for 25th, 50th and 75th quantiles of 75 mph for T2DM drivers. Insignificant differences were also captured at different quantiles for 75 mph and 65 mph in the T1DM category. This aspect, which couldn't be captured by the linear models (LMM) in the earlier study, was illuminated by our findings through the QR model [10].

TABLE V. ESTIMATED COEFFICIENTS OF QUANTILES OF DEVIATION FROM SPEED LIMIT DISTRIBUTIONS. SIGNIFICANCE (P-VALUE) LEVELS ARE GIVEN IN BRACKETS

| PSL (mph) | Variable | T1DM | | | | | T2DM | | | | |
|---|---|---|---|---|---|---|---|---|---|---|---|
| | | 25th | 50th | 75th | 85th | 90th | 25th | 50th | 75th | 85th | 90th |
| 75 | Intercept | 3.0417 (0.000) | 1.9557 (0.000) | 3.3923 (0.000) | 3.1707 (0.000) | 3.8070 (0.000) | -1.8590 (0.000) | -0.5468 (0.000) | **-0.1653 [d] (0.124)** | 0.3365 (0.003) | **-0.1245 [d] (0.433)** |
| | HC | 0.4438 (0.000) | 2.5374 (0.000) | 1.9944 (0.000) | 2.3947 (0.000) | 2.3068 (0.000) | 1.7731 (0.000) | 1.4585 (0.000) | 1.7913 (0.000) | 2.1892 (0.000) | 2.6750 (0.000) |
| | DHR | 2.6068 (0.000) | 1.6276 (0.000) | 0.4956 (0.003) | **-0.1056 [d] (0.663)** | 0.4270* (0.049) | -3.4876 (0.000) | -1.6745 (0.000) | -0.4960 (0.000) | 0.4821 (0.001) | 0.8495 (0.000) |
| | Peak-Hours | 0.7178 (0.007) | **0.0387 [d] (0.872)** | **0.2057 [d] (0.456)** | **-0.6434 [d] (0.103)** | -1.3434 (0.000) | -0.9297 (0.000) | -2.0890 (0.000) | -1.5370 (0.000) | -1.9724 (0.000) | -1.6970 (0.000) |
| | Weekday | -0.5562 (0.000) | -1.4480 (0.000) | -1.4866 (0.000) | -1.2433 (0.000) | -1.3660 (0.000) | -0.5680 (0.000) | 0.2379 (0.000) | 1.3461 (0.000) | 1.0870 (0.000) | 1.9854 (0.000) |
| | Female | -2.5675 (0.000) | 0.6887 (0.000) | 1.5246 (0.000) | 2.3866 (0.000) | 2.7588 (0.000) | 2.2421 (0.000) | 2.4923 (0.000) | 2.6224 (0.000) | 2.8130 (0.000) | 3.0507 (0.000) |
| 70 | Intercept | - | - | - | - | - | -11.7768 (0.000) | -3.82 (0.000) | -0.816 (0.001) | 1.065 (0.000) | 2.47 (0.000) |
| | HC | - | - | - | - | - | 6.225 (0.000) | 4.35 (0.000) | 3.783 (0.000) | 2.862 (0.000) | 2.034 (0.000) |
| | DHR | - | - | - | - | - | 5.289 (0.000) | 2.83 (0.000) | 2.350 (0.000) | 1.455 (0.000) | 1.210 (0.000) |
| | Peak-Hours | - | - | - | - | - | 4.154 (0.000) | **0.294 [d] (0.592)** | 0.769 (0.000) | 0.540 (0.001) | **-0.218 [d] (0.522)** |
| | Weekday | - | - | - | - | - | -1.59 (0.005) | -1.973 (0.000) | **-0.454 [d] (0.054)** | **-0.355 [d] (0.104)** | -0.529 (0.008) |
| | Female | - | - | - | - | - | **0.4067 [d] (0.396)** | 1.978 (0.000) | 0.850 (0.000) | **0.118 [d] (0.506)** | **0.1069 [d] (0.590)** |
| 65 | Intercept | -3.087 (0.000) | 0.6933 (0.001) | 2.7636 (0.000) | 4.488 (0.000) | 2.88 (0.000) | -5.76 (0.000) | -1.212 (0.000) | 1.066 (0.000) | 2.367 (0.000) | 3.988 (0.000) |
| | HC | -2.633 (0.000) | -2.23 (0.000) | -1.915 (0.000) | -1.919 (0.000) | **-0.126 [d] (0.761)** | 0.492 (0.006) | 1.365 (0.000) | 0.735 (0.000) | 0.402 (0.000) | **0.240 [d] (0.195)** |
| | DHR | **0.5122 [d] (0.104)** | **-0.0688 [d] (0.710)** | **0.0613 [d] (0.810)** | 2.537 (0.000) | 2.0687 (0.003) | 1.62 (0.000) | 1.378 (0.000) | 1.543 (0.000) | 1.778 (0.000) | 1.131 (0.003) |
| | Peak-Hours | 2.389 (0.000) | 0.4335 (0.032) | **0.2229 [d] (0.548)** | **0.158 [d] (0.903)** | **0.598 [d] (0.444)** | 2.40 (0.008) | 1.6474 (0.000) | -1.124 (0.001) | **-0.041 [d] (0.935)** | **0.580 [d] (0.46)** |
| | Weekday | **0.3456 [d] (0.097)** | 0.3535 (0.003) | 0.579 (0.003) | 0.923 (0.001) | 0.859* (0.045) | **-0.408 [d] (0.206)** | **-0.091 [d] (0.408)** | -0.187* (0.046) | **-0.078 [d] (0.651)** | **-0.067 [d] (0.804)** |
| | Female | -0.962 (0.005) | -2.467 (0.000) | -2.835 (0.000) | -2.930 (0.000) | **-1.264 [d] (0.108)** | **-0.298 [d] (0.112)** | 0.234 (0.000) | **-0.032 [b] (0.695)** | **-0.114 [d] (0.257)** | **0.047 [d] (0.807)** |

[d] Not significant at 95% level of confidence (boldface); * Significant at 95% level of confidence; - Not Applicable

- Among both T1DM and T2DM drivers, the HC drivers exhibited a notable increase in deviation from speed limit from the 75 mph posted speed limit compared to DN driving. Nevertheless, an opposing trend was observed for T1DM drivers at 65 mph.

The comparable speed adherence patterns observed in T1DM and T2DM drivers align with previous research indicating a connection between hyperglycemia and reduced cognitive performance in both type 1 and type 2 diabetes individuals [26]. Additionally, the impact of acute hyperglycemia on speed management, when contrasted with normal blood sugar levels (euglycemia), aligns with prior research indicating that individuals with T1DM impair both hazard perception and speed management abilities during DHR situations [10]. Nonetheless, variations in speed adherence behavior arise, likely attributed to the greater impact of hyperglycemia on driving behavior among T2DM subjects compared to T1DM. This aligns with a study showing that disruptive hyperglycemia was reported by 8% of T1DM drivers, in contrast to 40% of T2DM drivers [13].

## VI. CONCLUSIONS AND FUTURE AVENUES

This study provides novel results on how acute hyperglycemia impacts real-world driving in diabetes and uninterrupted traffic flow environments, like interstates, which may carry higher risks of fatal crashes.

Key results confirm previous reports linking deviation from speed limit to acute glucose episodes in T1DM and T2DM and advance understanding of hyperglycemia's role in diabetes speed control. Hyperglycemia may impact speed control in T2DM to a greater extent than T1DM. This aligns with newer research reporting that hyperglycemia in diabetes may impact driving more than previously estimated [14] along with the greater prevalence of hyperglycemia in T2DM compared to T1DM [13]. Results underscore the need for analytic approaches that better capture individual variability in driver safety and highlight the non-linear nature of these effects. An apparent factor contributing to the lesser use of quantile regression is its perceived interpretational complexity. Unlike linear regression, which provides a straightforward measure, quantile regression generates multiple coefficients that may not provide a cohesive and easily understandable picture.

Admittedly, this paper has limitations that we encourage investigators to address in future research. Notably, sample sizes for participants and glucose episodes were limited in this study. Video data, which was not used in this study, may improve understanding of how real-time traffic and roadway dynamics affect diabetes speed control. Future models may also further investigate other covariates (e.g., age, driving experience, and medication usage) on diabetes driver speed control. This may advance development of targeted, personalized interventions or strategies. An apparent factor contributing to the lesser use of quantile regression is its perceived interpretational complexity. Unlike linear regression, which provides a straightforward measure, quantile regression generates multiple coefficients that may not provide a cohesive and easily understandable picture.

Results shed light on specific driver behavior patterns, in context of individual variation, that may be used for future development of supportive, personalized in-vehicle technology. In-vehicle advanced driving systems show promise for aiding drivers with diabetes, and those with related disorders, in maintaining safe mobility. Patterns of driver speed control may show promise for identifying and intervening in at-risk driving situations.

ACKNOWLEDGMENT

The authors would like to acknowledge Toyota Collaborative Safety Research Center for funding this research and University of Nebraska Medical Center for their work recruiting and assessing this study's research participants.